\documentclass{iaus}
\usepackage{graphicx}

\title[H~II Region geometry, B fields, \& stellar feedback] 
{Magnetic fields, stellar feedback, and the geometry of H~II Regions}

\author[Gary J. Ferland]   
{Gary J. Ferland}

\affiliation{Department of Physics \& Astronomy,
University of Kentucky,
Lexington, KY 40506, USA
\break email: gjferland@gmail.com}

\pubyear{2009}
\volume{xxx}  
\pagerange{119--126}
\date{?? and in revised form ??}
\jname{Proceedings Title IAU Symposium}
\editors{A.C. Editor, B.D. Editor \& C.E. Editor, eds.}
\begin{document}

\maketitle

\begin{abstract}
Magnetic pressure has long been known to dominate over gas pressure
in atomic and molecular regions of the interstellar medium.
Here I review several recent observational studies of the relationships
between the H$^+$, H$^0$ and H$_2$ regions in
M42 (the Orion complex) and M17.
A simple picture results.
When stars form they push back surrounding material,
mainly through the outward momentum of starlight acting on grains,
and field lines are dragged with the gas due to flux freezing.
The magnetic field is compressed and the magnetic pressure increases
until it is able to resist further expansion
and the system comes into approximate magnetostatic equilibrium.
Magnetic field lines can be preferentially aligned perpendicular to
the long axis of quiescent cloud before stars form.
After star formation and pushback occurs ionized gas will be
constrained to flow along field lines and escape from the system
along directions perpendicular to the long axis.
The magnetic field may play other roles in the physics of
the H~II region and associated PDR.
Cosmic rays may be enhanced along with the field and provide additional
heating of atomic and molecular material.
Wave motions may be associated with the field and contribute a
component of turbulence to observed line profiles.
\end{abstract}

\firstsection 
\section{Introduction - The magnetic field of a quiescent cloud}

Magnetic fields play pivotal roles in star-forming environments.
Many aspects of this rich topic are covered in other papers
in this book, and the review by
Heiles \& Crutcher (2005) is essential reading.

The first of the many influences of the field is in the
formation of the molecular cloud itself.
The crucial physics is the coupling between the magnetic field
and even weakly ionized gas.
This so-called flux freezing means that there is a relationship,
set by the geometry of any expansion or contraction
that occurs, between the gas and field density.
This means that while gas is free to move along field lines,
gas motions perpendicular to the field will magnify or weaken the field.

Figure 1, taken from Heiles (1988),
shows the dark cloud L204.
The orientation of the magnetic field,
as deduced from starlight linear polarization,
is indicated by the black lines.
Field lines tend to lie perpendicular to the long axis of the filament.
The Pipe Nebula (Alves et al. 2008) is another example.
This geometry is not uncommon (Heiles \& Troland 2005).
One interpretation is that the field is strong enough to guide
contraction along field lines so that clouds tend to form
as sheets or filaments (Heitsch, Stone \& Hartmann 2009).
Gravitational contraction along the filament may further
strengthen the field.

\begin{figure}
 \includegraphics[scale=0.8,angle=90]{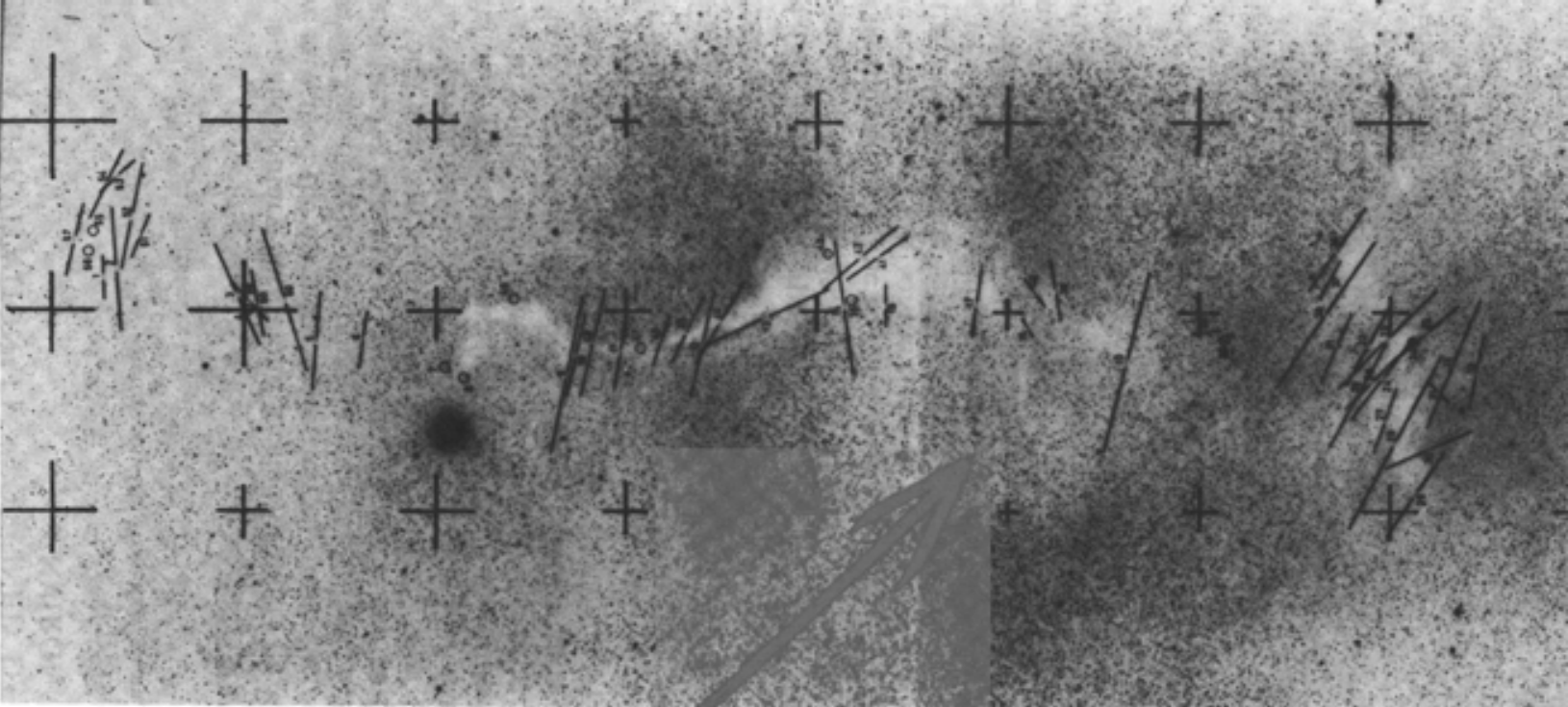}
  \caption{Figure 1, from Heiles (1988), showing
  the dark cloud Lynds 204.  Magnetic field lines,
  shown as black lines, line roughly orthogonal
  to the long axis of the filament.
  }\label{fig:Heiles88}
\end{figure}

\section{Orion - an active star-forming region}

\begin{figure}
 \includegraphics[scale=1.4,angle=270]{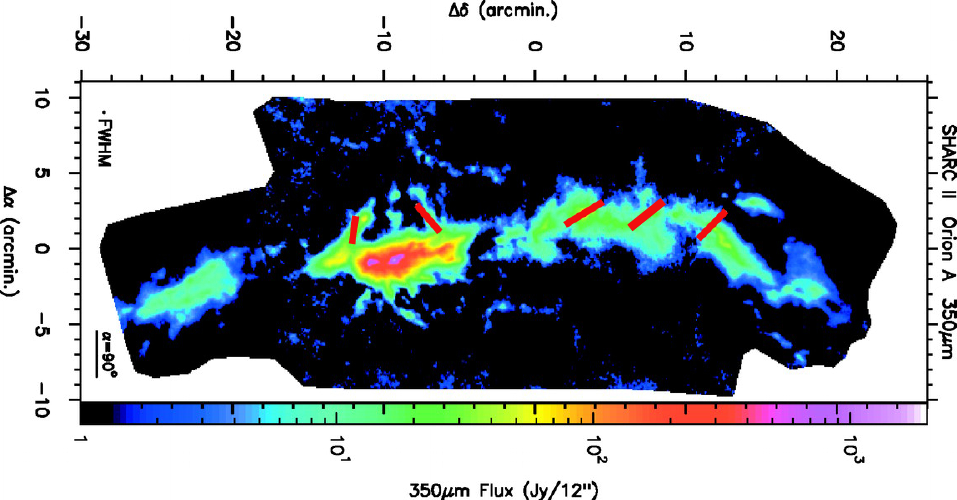}
  \caption{
  The Orion Molecular Cloud as imaged at $350 \mu$m
  by Houde et al (2004).
  The figure has been rotated so that north is to the right.
  The red lines show the direction of the magnetic field
  deduced from the linearly polarized dust emission.
  The Trapezium cluster is centered on the bright region
  to the left.
  }\label{fig:Houde04}
\end{figure}

The Orion complex is more complicated because of
the feedback associated with active star formation.
Figure \ref{fig:Houde04}, taken from Houde et al. (2004),
shows the geometry of the field as revealed by
linearly polarized dust emission.
The surface brightness of the $\lambda 350 \mu$m
thermal emission is shown by the colored scale.
The red lines indicate the deduced field direction.
The bright region to left of center is warm dust in molecular
material surrounding the Trapezium cluster.
Active star formation and associated starlight cause the dust to
be radiatively heated and glow brightly near the young stars.

The field lines tend to lie perpendicular to the long axis of the molecular cloud in the relatively quiescent northern regions.
This is reminiscent of the geometry of Lynds 204 shown in
Figure \ref{fig:Heiles88}.

\begin{figure}
\begin{centering}
 \includegraphics[scale=0.8]{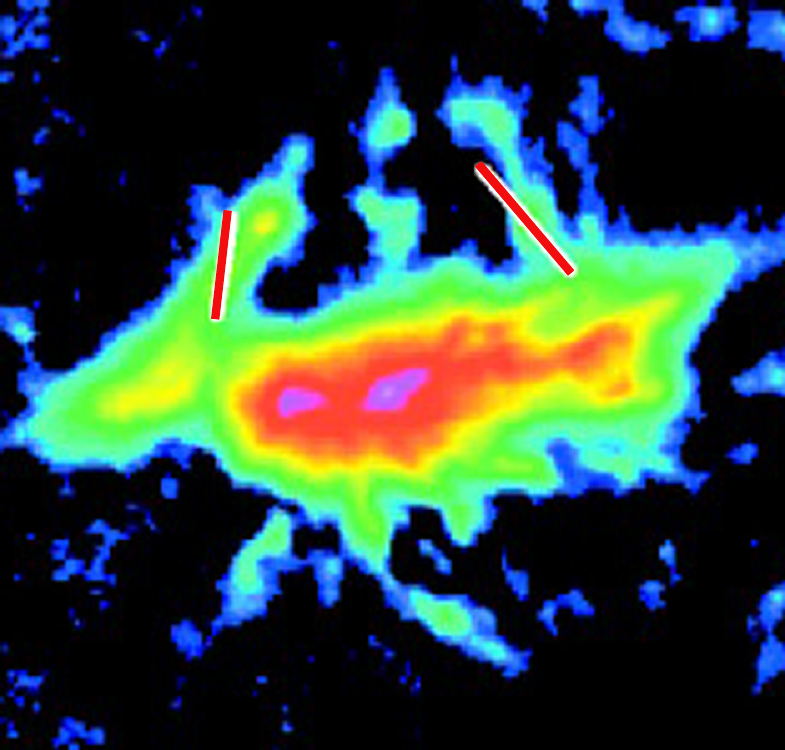}
  \caption{
  A zoom into Figure \ref{fig:Houde04} from Houde et al (2004)
  showing emission from the parent molecular cloud in
  the region around the Trapezium cluster.
  The orientation of the image is the same as in
  Figure \ref{fig:Houde04}.
  The orientation of the magnetic field is indicated by the red lines.
  }\label{fig:Houde04_trapezium}
\end{centering}
\end{figure}

The region surrounding the Trapezium is more complicated
due to the presence of massive stars and their associated
radiation pressure.
Figure \ref{fig:Houde04_trapezium} shows a zoom of the
region near the Trapezium.
Here the field lines tend to lie perpendicular to the line between the
point and the Trapezium.

\begin{figure}
\begin{centering}
 \includegraphics[scale=0.5]{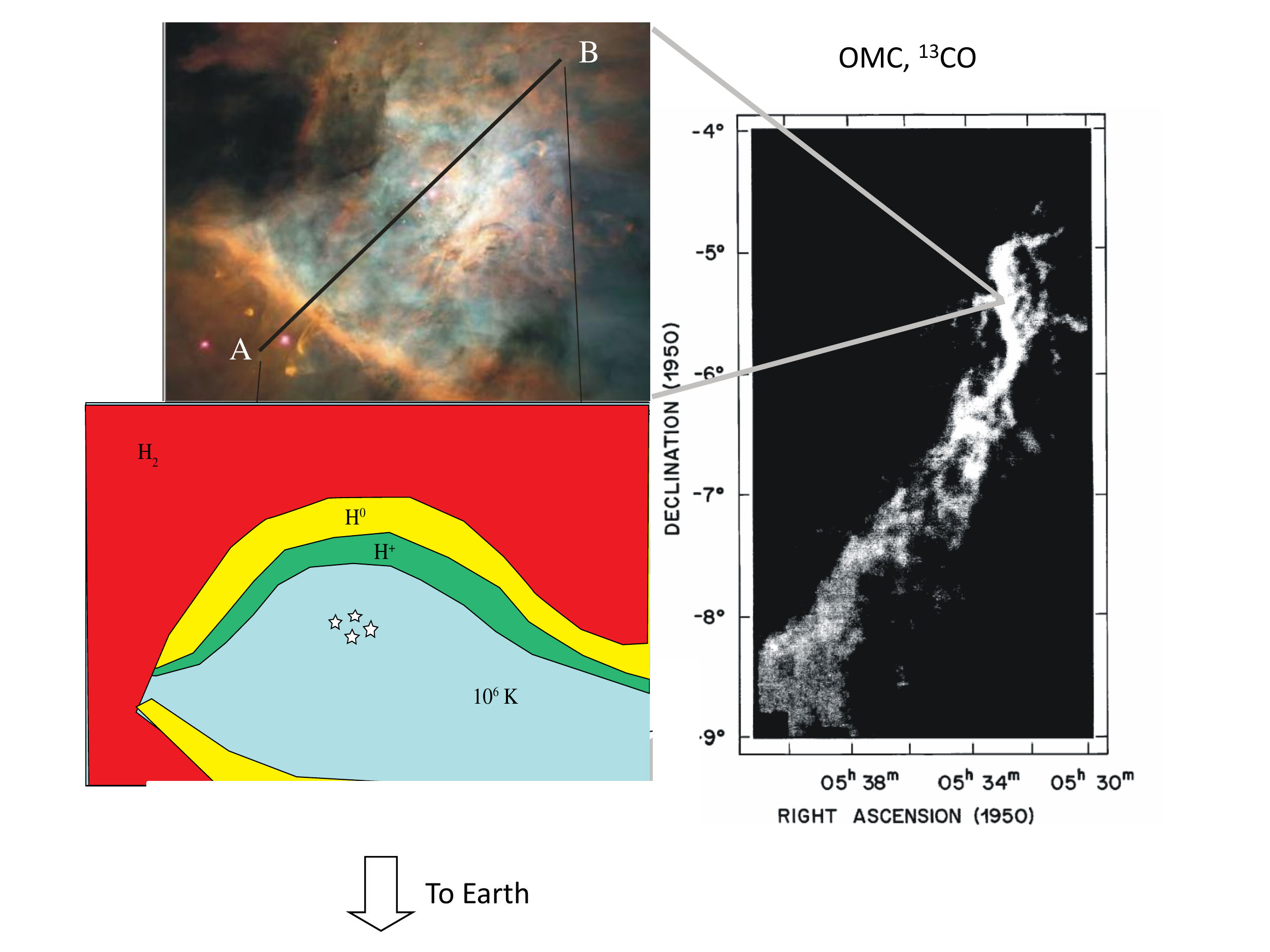}
 \end{centering}
  \caption{The geometry of the Orion H~II region, PDR,
  and molecular cloud.
  This Figure, adopted from Osterbrock \& Ferland (2006),
  shows a CO image of the molecular cloud to the right
  and a zoom into the HST image of the H~II region at upper left.
  The geometry of the cut shown as the black line from A to B
  in the HST image is shown in the lower left panel.
  The star cluster is surrounded by hot gas produced by stellar winds.
  The bright H~II region is mainly an ionized layer
  on the surface of the
  background molecular cloud.
  Much of the extinction visible in the HST image arises in
  Orion's ``Veil'', the layer of predominantly atomic gas that
  lies on the near side of the star cluster.
  The Veil is the region where H~I 21 cm circular polarization measures
  the line-of-sight magnetic field.
  }\label{fig:orion_structure}
\end{figure}

This review centers on the relationships between stars and
the surrounding H$^+$, H$^0$, and H$_2$ layers, the
so-called H~II region, PDR, and molecular cloud.
Figure \ref{fig:orion_structure} shows the geometry of the regions
near the young star cluster and bright H~II region in
the Orion complex.
This is adopted from Osterbrock \& Ferland (2006).

Feedback from Trapezium cluster has strongly affected
the cloud geometry.
Stellar winds and associated shocks produce a bubble of hot gas
that has recently been detected in the X-rays (G{\"u}del et al. 2007).
A combination of thermal gas pressure and starlight momentum pushes
cooler gas away from the star cluster and results in the blister
geometry shown in Figure \ref{fig:orion_structure}.

Most of the extinction seen in the HST image of the Orion H~II Region
arises in the Veil, the layer of predominantly atomic gas that lies
on this side of the hot bubble (O'Dell 2001a).
The Veil has been extensively studied at H~I 21 cm.
The thermal continuum emission produced by the H$^+$ region
is used to probe the Veil, where
atomic gas produces a 21 cm absorption line.
Zeeman measurements of the line of sight magnetic field show it to
be surprisingly strong, approaching 50 $\mu$G, roughly 1 dex
stronger than the field in the diffuse ISM
(Troland et al. 1989).
Why is the field so strong?

Abel et al. (2004; 2006) combined optical and UV measurements
of absorption lines formed in the Veil to derive its
density, kinetic temperature (and so its gas pressure) and its
distance from the Trapezium.
They found that magnetic pressure greatly exceeded the gas pressure,
as is typical of the ISM, and that the magnetic pressure exceeded
even the turbulent pressure in one of the two Veil components.
The Veil is a thin sheet which we view roughly face on.

\section{M17 and magnetostatic equilibrium}

The M17 star-forming region is much larger and more luminous
than Orion but also much further away.
Similar Zeeman polarization measurements of the magnetic field in
the atomic hydrogen region have been performed
(Brogan \& Troland 2001) and an even stronger field, approaching
700 $\mu$G, was found.

\begin{figure}
\begin{centering}
 \includegraphics[scale=0.5]{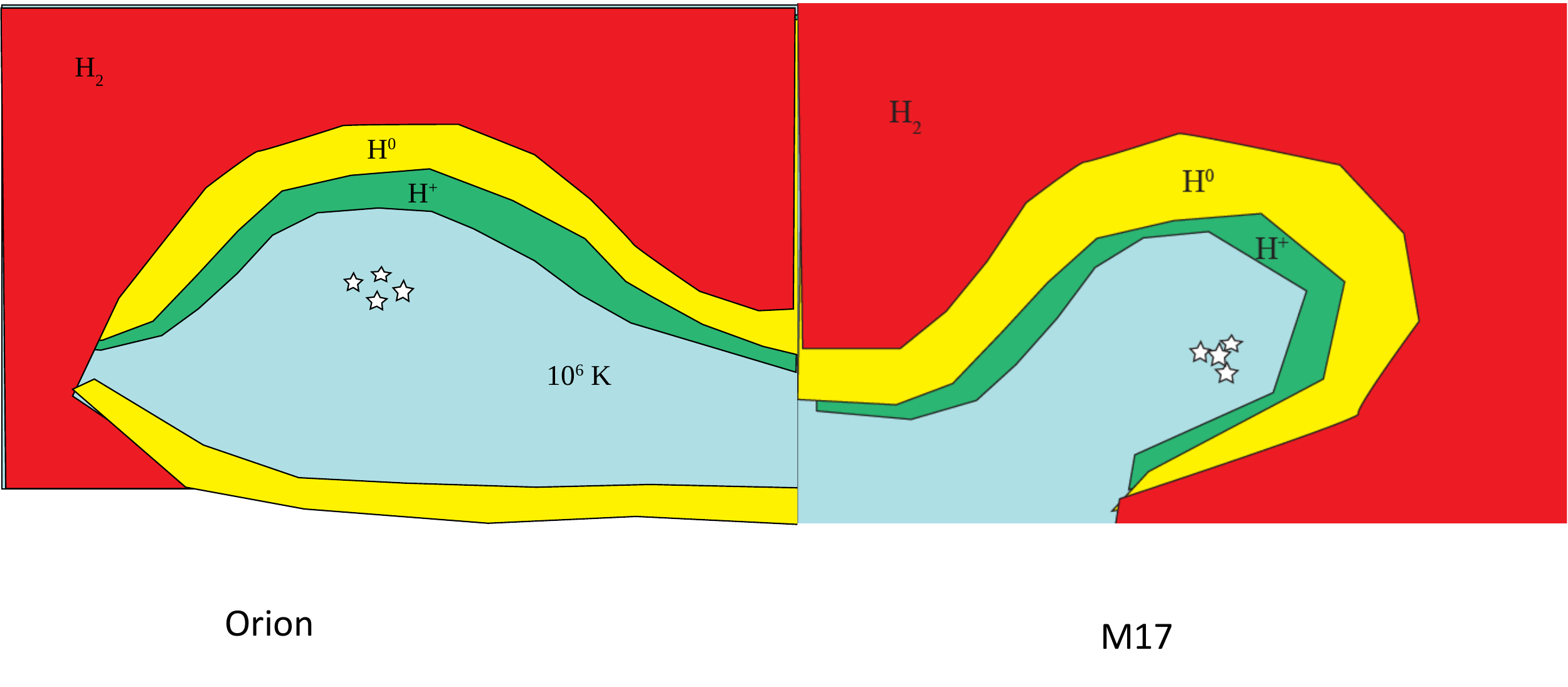}
 \end{centering}
  \caption{
A comparison of the geometry of Orion (left) and M17 (right).
This is not drawn to scale - M17 is roughly 1 dex larger than Orion.
We view both geometries from the bottom looking up.
The atomic layer in front of the star cluster and H~II region is
translucent in Orion.
In M17 the layer has a large extinction and has atomic, ionized,
and molecular constituents.
The Zeeman magnetic field measurements use the emission of the
H$^+$ layer as the continuum source.
The radio absorption measurements probe the line of sight component of
the magnetic field in the H$^0$ layer between the observer and
the H$^+$ region.
  }\label{fig:Orion_M17}
\end{figure}

Pellegrini et al (2007 hereafter P07)
combined a broad range of spectral observations to make
a coherent picture of the geometry of M17.
M17 has an overall geometry that is similar to Orion, but viewed
from a a different angle, as shown in Figure \ref{fig:Orion_M17}.
Some of these ideas are further outlined in Ferland (2008).
The ionizing star cluster and the intrinsically brightest part of the
H~II region are hidden behind a layer of atomic and molecular gas.
The Zeeman magnetic field is stronger in the regions to the right
(West) of the
star cluster then along directions more towards the cluster.

P07 showed that the H$^+$ H$^0$ and H$_2$ layers were in
a state of quasi-magnetostatic equilibrium.
The outward force of starlight, mainly ionizing radiation acting
on gas and dust, is resisted by the magnetic pressure in deeper
regions of the cloud.
The picture they proposed is that the combination of thermal
gas pressure from the hot bubble and radiation pressure due to
starlight has pushed back surrounding gas, strengthening the
magnetic field,
until the magnetic pressure could resist further compression.

The gas pressure in the hot wind-blown bubble
is close to the pressure near the illuminated face of the H~II Region.
The absorption of the outward-flowing starlight pushes the H~II
region away from the cluster increasing the gas density
and magnetic pressure.
The magnetic pressure increases as the square of the field, so
for many geometries the magnetic pressure will increase faster than
the gas pressure.
Most starlight is absorbed by the outermost parts of the H$_2$ layer
due to the large dust extinction.
Magnetic pressure dominates at this point so the total
luminosity of the star cluster and the magnetic field at this point
are related.

P07 gave a simple relationship between the total luminosity of the star
cluster and the magnetic pressure in deeper regions of the cloud.
In this picture the strong magnetic fields associated with active
regions of star formation are directly related to the luminosity
of the central stars.
This provides a natural explanation for why strong fields are found
near star-forming regions.

\section{The magnetic field and the geometry of the M17 H~II region}

The linear polarization measurements (Figures 1 \& 2)
suggest that the magnetic
field may be roughly perpendicular to the axis of a filamentary
molecular cloud.
What happens when star formation occurs in such a geometry?
Two forces act to guide the resulting expansion of the ionized gas.

\begin{figure}
 \includegraphics[scale=1.4]{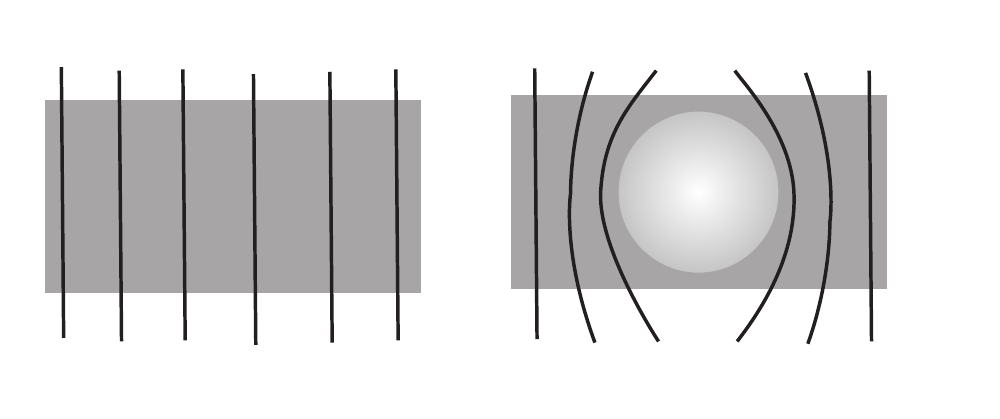}
  \caption{
A cartoon showing the expansion of an H~II region into
an ordered magnetic field.
The left panel shows a quiescent cloud with the field lines
orthogonal to the axis of the cloud.
The right panel shows the expanded ionized gas with associated
changes in the field geometry.
  }\label{fig:FieldLines}
\end{figure}

The first is the effects of an ordered magnetic field as shown
in Figure \ref{fig:FieldLines}.
The left panel shows a segment of a filamentary molecular cloud
with an ordered field.
The right panel shows the expanded hot bubble and compressed
field lines.
The magnetic field increases perpendicular to the ``equator'' of the
cloud.
Expansion can be halted by the field in this direction.
There is no increase along the ``poles'' and gas may be free
to move in this direction.

\begin{figure}
 \includegraphics[scale=0.5,angle=270]{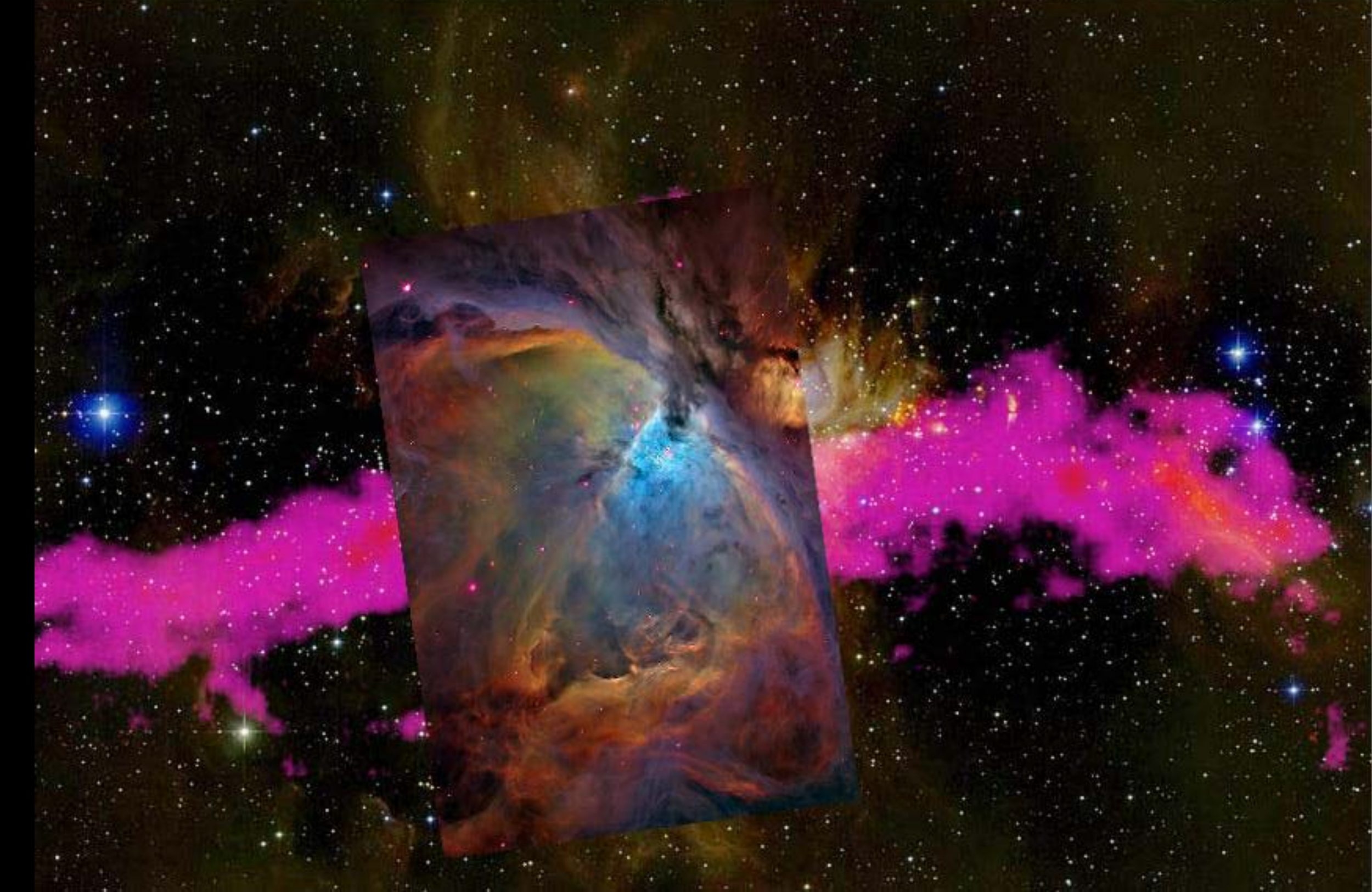}
  \caption{
William J. Henney's composite of an optical
image of the Orion H II region in the center and
a radio image showing the distribution of carbon
monoxide molecules in red.
Optical image obtained as part of the Two
Micron All Sky Survey (2MASS), a joint project
of the University of Massachusetts and the Infrared
Processing and Analysis Center/California
Institute of Technology, funded by the National
Aeronautics and Space Administration and the
National Science Foundation.  The CO image is
derived from Plume et al.(2000).
The x-ray emission, which is not shown, occurs below the
``fan'' of optical emission which points below the
molecular cloud.
  }\label{fig:Henney_Orion}
\end{figure}

Figure \ref{fig:Henney_Orion} shows a composite image
of the Orion molecular cloud, star cluster, and H~II Region.
An image with higher resolution data is given in Henney (2008).
Figure \ref{fig:Henney_Orion} shows that the visible
H~II region opens in the direction below the molecular cloud
in the image.
This is the direction where the X-ray emission discovered by
G{\"u}del et al. (2008) occurs and is the likely direction of outflow
of the hot gas.
The geometry of the magnetic field shown in Figure \ref{fig:FieldLines}
may account for the direction of the extended optical emission
since gas will only be free to move
in directions perpendicular to the filament.

Linear polarization shows the direction of the magnetic field.
Figure \ref{fig:Houde04_trapezium}, which is a zoom into a portion
of the larger map given in Figure \ref{fig:Houde04},
suggests that the field lines near the Trapezium are roughly
orthogonal to the direction towards the star cluster.
This field geometry is consistent with the picture
shown in Figure \ref{fig:FieldLines}.

Zeeman circular polarization measurements detect the line of sight
component of the magnetic field.
For the geometry shown in Figure \ref{fig:FieldLines}
observations in the direction of the star cluster
along the equator of the slab will detect
the smallest field.
The line-of-sight field should increase as the telescope
beam moves away from the cluster and the impact parameter
of the passing ray increases.
This is the general sense of the measured magnetic field of Orion
and M17.
The magnetic field tends to be stronger in directions that do not
center on the star cluster.
But the field does not go to zero in the direction of the cluster
in Orion.
(Measurements do not exist in this direction in M17.)
This may indicate that the field has a disorganized
component in addition to the ordered component detected by
linear polarization.

The mass shaping due to the large mass of the
molecular cloud is a second effect influencing the
expansion of the H~II region.
In both Orion and M17 star formation appears to have occurred
near a surface of the molecular cloud
and expanding x-ray emitting gas is freely expanding out into
the general ISM along along an open path.
There is relatively little material along our line of sight to
the Trapezium while the main outflow appears to be towards
the Southwest (towards the bottom in Figure \ref{fig:Henney_Orion}).
In M17 there is a large column density of material along our
line of sight towards
the cluster, which is optically obscured, while the
outflowing hot gas is directed towards the East
(to the left in the right panel of Figure \ref{fig:Orion_M17}).
In both cases a large mass of molecular gas is present in some
directions but not in others.
This mass shaping must also affect the geometry of the ionized bubble.

\section{Are cosmic rays amplified as well?}

Cosmic rays are in approximate energy equipartition
with the magnetic field in the diffuse ISM (Webber 1998).
Energy equipartition is usually assumed when observations of radio synchrotron
emission are interpreted in terms of a cosmic ray density
or magnetic field, as reviewed
in other papers in this volume.
Such equipartition does not occur because of any direct microphysical
coupling between the energy reservoirs in the environment.
Rather it is a minimum energy configuration that can be established
if the system exists for a long enough time to have become relaxed.
The equipartition value of the cosmic ray density does provide a
reference point to which I will come back below.

Low-energy cosmic ray electrons are strongly trapped
by the magnetic field because
their gyro radius is much smaller than the physical scales of
the molecular cloud or H~II region.
The compression of field lines shown in Figure \ref{fig:FieldLines}
will also increase the cosmic ray density.
Cosmic rays will be free to drift along field lines unless their pitch angle
is perfectly aligned with the field.
They should drift out of the environment relatively quickly.
This drift would be inhibited if the field has a disorganized component.
Low-energy cosmic rays can be
locally enhanced by MHD waves (Padoan \& Scalo 2005)
so spatial variations are expected.
This is clearly a complicated topic.

The galactic cosmic ray density is usually derived from observations
of the ion chemistry (Spitzer 1978).
Recent observations of the fraction of gas in H$_3^+$ along a
number of sight lines suggest that the galactic
background cosmic ray ionization rate has been underestimated
by about 1 dex, and that this rate varies from one sight line
to another (Indriolo et al. 2007).
This shows that the cosmic ray density of even quiescent regions
is open to question.

Cosmic rays both heat and ionize gas.
The proportion of energy that goes into heating or ionization is
determined by the degree of ionization of the gas
(Osterbrock \& Ferland 2006).
Heating by cosmic rays is not insignificant in atomic regions
even with the old lower background rate (Tielens \& Hollenbach 1985).
If the cosmic ray rates are significantly higher than previously
thought then they will be even more important in heating atomic and
molecular regions.

Giammanco \& Beckman (2005) proposed that a high flux of
cosmic rays may explain the ``temperature fluctuations'' problem
in H~II regions.
The basic conundrum is that different measures of the kinetic
temperature in the ionized gas do not agree with one another.
The disagreement is systematic and suggests that a range of
kinetic temperatures is present.
It is hard to understand how this can occur due to the
rapid increase of the gas cooling function with increasing
temperature (Ferland 2003) and the fact that a Boltzmann distribution
is established so quickly in an ionized gas (Spitzer 1978).
Giammanco \& Beckman argue that the supplemental heating
provided by enhanced cosmic rays would produce the observed effects.

Orion's ``Bar'', a linear feature obvious on most optical images
of the H~II region (O'Dell 2001a),
is thought to be an escarpment on the surface of the background
molecular cloud.
Ionizing radiation from the Trapezium strikes the Bar in such
a way that we can observe the transition between
H$^+$ H$^0$ and H$_2$ regions nearly edge on.
Because of its special geometry, the Bar has become a decisive
test for the physics of the interface between molecular and ionized regions.

Pellegrini et al. (2008) and Shaw et al. (2009) studied
spectral variations across the Bar with the goal of reproducing
the observed emission profiles of various tracers of ionized,
atomic, and molecular gas.
They found that cosmic rays in equipartition with the
observed strong magnetic field
reproduced the observations.
P08 also required an enhanced flux of cosmic rays to account for
the atomic and molecular emission in Orion.

\section{Magnetic fields and non-thermal line widths}

ISM spectral lines, both emission and absorption, are usually
found to have widths too large to be due to purely thermal motions.
These non-thermal line widths were
often interpreted as a form of non-dissipative wave motion
with the kinetic and magnetic energies in rough energy equipartition.
Numerical MHD simulations suggest that such waves should be damped, which
then requires that some energy source drive them, but in any case
the turbulently broadened lines can be taken as empirically motivated.
Heiles \& Crutcher (2005) review the situation.

It is successively more difficult to measure non-thermal line widths
in PDRs and H~II regions than in molecular clouds
due to the increasing temperature and decreasing mass per particle.
A given turbulent motion, which might be associated with a
particular magnetic field, is a smaller fraction of the thermal line
width in hot ionized gas than it would be in a cold molecular medium.
Non-thermal line widths are seen in the spectra of
the H~II regions and PDRs.
PDR lines have long been known to be turbulently broadened
(Tielens \& Hollenbach 1985) and
Roshi (2007) has recently suggested that non-thermal line widths of
carbon radio recombination lines, which should form
in the PDR, may be associated with MHD waves.

Optical emission lines from the H~II region also have larger-than-expected
line widths.
These widths are observed to scale with the luminosity of the
star cluster and have been
studied extensively because of their possible use as a standard candle
(Melnick et al. 1988).
For low-luminosity objects like Orion the observation is difficult because
the expansion of the ionized gas away from the molecular cloud,
which occurs at roughly the speed of sound in the H~II region ($\sim 10$
km s$^{-1}$), is similar to the turbulent speed.
However, very careful studies have been done
(O'Dell et al 2005; Garc\'ia-D\'iaz et a. 2008) and find a significant
component of turbulence.
There is no agreed-upon model for the origin of the turbulence seen in
the optical lines but MHD waves are a possibility
(Ferland 2001; O'Dell 2001a, 2001b).
This is important because the turbulent energy is a
significant part of the energy budget in Orion.
If this turbulence is dissipative, as suggested by
numerical MHD simulations, then it would act to heat the gas
(Pan \& Padoan 2008).

Line widths in excess of
100 km s$^{-1}$ can be found in luminous extragalactic H~II regions.
Beckman \& Rela{\~n}o (2004) argue that these line widths
may be due to MHD effects associated with the magnetic field.
This would imply that relationships between magnetic field,
turbulent, and gravitational energies extend over many orders of
magnitude of mass.
Again, if the waves are dissipative then,
depending on the damping timescale,
they may constitute a significant heating source.

\section{Conclusions}\label{sec:conclusions}

\begin{enumerate}
\renewcommand{\theenumi}{\arabic{enumi}}

\item
A number of studies of the H$^+$, H$^0$ / H$_2$ regions of active star-forming regions,
chosen to have high-quality Zeeman 21 cm detections
of the magnetic field in the H$^0$ region,
have been conducted.
In both Orion's Veil and the H$^0$ region along the line of sight
to the H$^+$ region in M17 the magnetic pressure is much
greater than the gas pressure, as is
commonly found in the ISM.
The magnetic pressure exceeds the turbulent pressure in one
component of the Veil.

\item
The strongest magnetic fields in diffuse gas are found
near regions of active star formation.
The field in the atomic hydrogen region of M17,
$\sim 700 \mu$G, is among the strongest observed.
This corresponds to a magnetic pressure roughly 3 dex larger
than is found in the diffuse ISM.

\item
Simulations of the emission-line spectra of
the H$^+$, H$^0$ / H$_2$ regions in M17 which reproduce
the observed magnetic field show that the geometry
is approximately in magnetostatic equilibrium.
The outward radiation pressure due to starlight is balanced
by the magnetic pressure in well-shielded regions of the cloud.
This accounts for the large field that is observed.

\item
Linear polarization studies suggest that the field is
often oriented perpendicular to the long axis of filamentary clouds.
This may be the result of the magnetic field guiding the formation
of the cloud or of the subsequent gravitational contraction along
the filament (Heitsch et al. 2009).

\item
Cosmic rays are found to be in energy equipartition with the
magnetic field in the diffuse ISM, and equipartition is
often assumed to hold for synchrotron emitting regions.
Cosmic rays both heat and ionize atomic and molecular gas.
If cosmic rays are enhanced along with the field then they would
constitute an important heating source for atomic and
molecular regions.

\item
Non-thermal line widths are often found in
emission line spectra of star-forming regions.
Several recent studies have suggested that these may be
due to MHD waves associated with the magnetic field.
If the waves are dissipative then this would constitute a
gas heating mechanism.

\end{enumerate}

\begin{acknowledgments}
Support by the NSF (AST 0607028) and NASA (ATFP07-0124)
is gratefully acknowledged.
I thank Carl Heiles and Will Henney for providing figures
and Bob O'Dell and Tom Troland for comments.
\end{acknowledgments}

\end{document}